\documentclass[aps,preprint,nofootinbib,showpacs]{revtex4}
\usepackage[dvips]{graphicx,color}
\usepackage{amsmath}
\newcommand{\be}{\begin{equation}}
\newcommand{\ee}{\end{equation}}
\newcommand{\ba}{\begin{eqnarray}}
\newcommand{\beq}{\begin{equation}}
\newcommand{\eeq}{\end{equation}}
\newcommand{\ea}{\end{eqnarray}}

\newcommand{\eV}{\text{eV}}
\newcommand{\MeV}{\text{MeV}}
\newcommand{\GeV}{\text{GeV}}
\newcommand{\TeV}{\text{TeV}}

\newcommand{\BR}{\text{BR}}

\newcommand{\meee}{{\mu \to \bar{e}ee}}

\newcommand{\mll}{{m_{\ell\ell^\prime}^{}}}
\newcommand{\hll}{{h_{\ell\ell^\prime}^{}}}
\newcommand{\qqHppHmm}{{q\overline{q}\to \gamma^*,Z^* \to H^{++}H^{--}}}
\newcommand{\ppHppHmm}{{pp\to \gamma^*,Z^* \to H^{++}H^{--}}}

\newcommand{\ppHpHm}{{pp\to \gamma^*,Z^* \to H^{+}H^{-}}}
\newcommand{\qqHpmpmHmp}{{q^\prime\overline{q}\to W^* \to H^{\pm\pm}H^\mp}}
\newcommand{\ppHpmpmHmp}{{pp\to W^* \to H^{\pm\pm}H^\mp}}
\newcommand{\ppHppHm}{{pp\to W^* \to H^{++}H^-}}
\newcommand{\ppHmmHp}{{pp\to W^* \to H^{--}H^+}}
\newcommand{\ppHpmH}{{pp\to W^* \to H^\pm H^0}}
\newcommand{\ppHpH}{{pp\to W^* \to H^+H^0}}
\newcommand{\ppHmH}{{pp\to W^* \to H^-H^0}}
\newcommand{\ppHpmA}{{pp\to W^* \to H^\pm A^0}}
\newcommand{\ppHA}{{pp\to Z^* \to H^0 A^0}}
\newcommand{\Hpmlpmnu}{{H^\pm \to \ell^\pm \nu_{\ell^\prime}^{}}}

\def\beqa{\begin{eqnarray}}
\def\eeqa{\end{eqnarray}}

\def\bea{\begin{eqnarray}}
\def\eea{\end{eqnarray}}

\def\err#1#2{\lower2pt\hbox{ $\stackrel{\scriptstyle +#1}{\scriptstyle -#2}$}}
\def\ga{\mathrel{\raise.3ex\hbox{$>$\kern-.75em\lower1ex\hbox{$\sim$}}}}
\def\la{\mathrel{\raise.3ex\hbox{$<$\kern-.75em\lower1ex\hbox{$\sim$}}}}
\def\bmaT{\left(\begin{array}{ccc}}
\def\emaT{\end{array}\right)}
\def\bma{\left( \begin{array} }
\def\ema{\end{array} \right)}
\def\gsim{~{\rlap{\lower 3.5pt\hbox{$\mathchar\sim$}}\raise 1pt\hbox{$>$}}\,}
\def\lsim{~{\rlap{\lower 3.5pt\hbox{$\mathchar\sim$}}\raise 1pt\hbox{$<$}}\,}

\begin{document}

\preprint{
\vbox{%
\hbox{SHEP-11-09}
}}
\title{\boldmath Production of doubly charged scalars from the decay 
of \\singly charged scalars in the Higgs Triplet Model\unboldmath} 
\author{A.G. Akeroyd}
\email{a.g.akeroyd@soton.ac.uk}
\affiliation{NExT Institute and School of Physics and Astronomy, University of Southampton \\
Highfield, Southampton SO17 1BJ, United Kingdom}
\author{Hiroaki Sugiyama}
\email{hiroaki@fc.ritsumei.ac.jp}
\affiliation{Department of Physics,
Ritsumeikan University, Kusatsu, Shiga 525-8577, Japan}

\date{\today}
\begin{abstract}
 The existence of doubly charged Higgs bosons ($H^{\pm\pm}$)
is a distinctive feature of the Higgs Triplet Model (HTM), in which neutrinos
obtain tree-level masses from the vacuum expectation value of a 
neutral scalar in a triplet representation of $SU(2)_L$.
We point out that a large branching ratio for the decay of a singly charged Higgs boson to
a doubly charged Higgs boson via $H^\pm\to H^{\pm\pm}W^*$
is possible in a sizeable parameter space of the HTM\@.
From the production mechanism $q'\overline q\to W^* \to H^{\pm\pm}H^\mp$ the above decay
mode would give rise to pair production of $H^{\pm\pm}$, with a cross section 
which can be comparable to that of
the standard pair-production mechanism $\qqHppHmm$.
We suggest that the presence of a sizeable
branching ratio for $H^\pm\to H^{\pm\pm}W^*$ could significantly enhance the detection prospects of $H^{\pm\pm}$ 
in the four-lepton channel.
Moreover, the decays $H^0\to H^\pm W^*$ and $A^0\to H^\pm W^*$
from production of the neutral triplet scalars $H^0$ and $A^0$ would also provide an additional source
of $H^\pm$, which can subsequently decay to $H^{\pm\pm}$.

\end{abstract}
\maketitle


\section{Introduction} 
The established evidence that neutrinos oscillate
and possess small masses~\cite{Fukuda:1998mi}
necessitates physics beyond the Standard Model
(SM), which could 
manifest itself at the CERN Large Hadron Collider (LHC) 
and/or in low energy experiments which 
search for the lepton flavour violation~\cite{Kuno:1999jp}.
Consequently, models of neutrino mass generation which can be probed
at present and forthcoming experiments are of great phenomenological
interest. 

Neutrinos may obtain mass via the vacuum expectation value
(vev) of a neutral Higgs boson in an isospin triplet
representation~\cite{Konetschny:1977bn, Mohapatra:1979ia, Magg:1980ut,Schechter:1980gr,Cheng:1980qt}.
A particularly simple implementation of this
mechanism of neutrino mass generation is the 
``Higgs Triplet Model'' (HTM)
in which the SM Lagrangian is augmented solely by $\Delta$
which is a $SU(2)_L$ triplet of scalar particles
with hypercharge $Y=2$~\cite{Konetschny:1977bn, Schechter:1980gr,Cheng:1980qt}.
 In the HTM,
the Majorana neutrino mass matrix $m_{\ell\ell^\prime}^{}$
($\ell,\ell^\prime=e,\mu,\tau$)
is given by the product of
a triplet Yukawa coupling matrix $\hll$
and a triplet vev ($v_\Delta$).
 Consequently,
the direct connection between $\hll$ and $m_{\ell\ell^\prime}^{}$
gives rise to phenomenological predictions
for processes which depend on $\hll$
because $\mll$ has been restricted well by neutrino oscillation measurements%
~\cite{Fukuda:1998mi,solar,atm,acc,Apollonio:2002gd,:2008ee}.
A distinctive signal of the HTM would be the observation of doubly charged Higgs bosons ($H^{\pm\pm}$)
whose mass ($m_{H^{\pm\pm}}$) may be of the order of the electroweak scale.
 Such particles 
can be produced with sizeable rates at hadron colliders in the
processes $\qqHppHmm$~\cite{Barger:1982cy, Gunion:1989in, Muhlleitner:2003me, Han:2007bk, Huitu:1996su}
and $\qqHpmpmHmp$~\cite{Barger:1982cy, Dion:1998pw, Akeroyd:2005gt}.
The first searches for $H^{\pm\pm}$ at a hadron collider 
were carried out at the Fermilab Tevatron, assuming the 
production channel
$\qqHppHmm$ and decay $H^{\pm\pm}\to \ell^\pm{\ell^\prime}^\pm$. 
The mass limits $m_{H^{\pm\pm}}> 110\to 150\,\GeV$~\cite{Acosta:2004uj,Aaltonen:2008ip} were derived, with the
strongest limits being for $\ell=e,\mu$~\cite{Acosta:2004uj}.
The branching ratios (BRs) for 
$H^{\pm\pm}\to \ell^\pm{\ell^\prime}^\pm$ depend on $\hll$
and are predicted in the HTM in terms of the parameters
of the neutrino mass matrix~\cite{Akeroyd:2005gt, Ma:2000wp, Chun:2003ej}.
Detailed quantitative studies of
BR($H^{\pm\pm}\to \ell^\pm{\ell^\prime}^\pm$) in the HTM have been performed in
\cite{Garayoa:2007fw,Akeroyd:2007zv,Kadastik:2007yd,Perez:2008ha} 
with particular emphasis  
given to their sensitivity to the Majorana phases and 
the absolute neutrino mass i.e.\ parameters which cannot be 
probed in neutrino oscillation experiments. 
 A study on the relation between BR($H^{\pm\pm}\to \ell^\pm{\ell^\prime}^\pm$)
and the neutrinoless double beta decay can be seen in \cite{Petcov:2009zr}.
Simulations of the detection prospects of $H^{\pm\pm}$ at the LHC 
with $\sqrt s=14\,\TeV$ previously focussed on
$\qqHppHmm$ only~\cite{Azuelos:2005uc},
but recent studies now include
the mechanism $\qqHpmpmHmp$~\cite{Perez:2008ha, delAguila:2008cj, Akeroyd:2010ip}. 
The first search
for $H^{\pm\pm}$ at the LHC with $\sqrt s=7\,\TeV$~\cite{CMS-search}
has recently been performed for both
production mechanisms $\qqHppHmm$ and $\qqHpmpmHmp$,
for the decay channels $H^{\pm\pm}\to \ell^\pm{\ell^\prime}^\pm$
and $\Hpmlpmnu$.

In phenomenological studies of the HTM, for simplicity it is sometimes assumed that $H^{\pm\pm}$ and $H^{\pm}$ are degenerate, with
a mass $M$ which arises from a bilinear term
$M^2{\rm Tr}(\Delta^\dagger \Delta)$ in the scalar potential.
In this scenario the only possible decay channels for $H^{\pm\pm}$ are
$H^{\pm\pm}\to \ell^\pm{\ell^\prime}^\pm$ and $H^{\pm\pm}\to W^\pm W^\pm$, and the branching ratios are
determined by the magnitude of $v_\Delta$.
However, quartic terms in the scalar potential
break the degeneracy of $H^{\pm\pm}$ and $H^{\pm}$, and induce
a mass splitting $\Delta M\equiv m_{H^{\pm\pm}}-m_{H^{\pm}}$,
which can be of either sign.
If $m_{H^{\pm\pm}}> m_{H^{\pm}}$ then a new decay channel becomes available for
$H^{\pm\pm}$, namely $H^{\pm\pm}\to H^{\pm}W^*$. 
Some attention has been given to the decay $H^{\pm\pm}\to H^{\pm}W^*$, and it has been
shown that it can be the dominant channel over a wide range of 
values of $\Delta M$ and $v_\Delta$~\cite{Chakrabarti:1998qy,Chun:2003ej,Akeroyd:2005gt,Perez:2008ha}, even for
$\Delta M \ll m_W$.

Another scenario is the case of $m_{H^{\pm}}> m_{H^{\pm\pm}}$, which would give rise to a new decay channel for the singly charged scalar, 
namely $H^\pm\to H^{\pm\pm}W^*$. This possibility has been mentioned in the context of the HTM in \cite{Chun:2003ej} only. We will
perform the first study of the magnitude of its branching ratio, as well as quantify its contribution to the production of
$H^{\pm\pm}$ at the LHC\@.%
\footnote{The decay $H^{\pm}\to H^{\pm\pm}W^*$ has also been briefly mentioned
in \cite{Babu:2009aq} in the context of a model with an isospin 3/2 multiplet with 
hypercharge $Y=3$, which also includes triply charged Higgs bosons.} The decay rate for $H^\pm\to H^{\pm\pm}W^*$ is easily obtained from
the corresponding expression for the decay rate for $H^{\pm\pm}\to H^{\pm}W^*$, and thus one expects that $H^\pm\to H^{\pm\pm}W^*$ will be
sizeable over a wide range of values of $\Delta M$ and $v_\Delta$.
We point out for the first time that the decay $H^\pm\to H^{\pm\pm}W^*$ would give rise to an alternative way to produce
$H^{\pm\pm}$ in pairs  ($H^{++}H^{--}$), namely by the production mechanism
$\qqHpmpmHmp$ followed by $H^\mp\to H^{\pm\pm}W^*$.
Production of $H^{++}H^{--}$ can give rise to a distinctive signature of four leptons ($\ell^+\ell^+\ell^-\ell^-$), and 
simulations and
searches of this channel currently only assume production via the process
$\qqHppHmm$.

Our work is organised as follows. In section~II we describe the theoretical structure of the HTM\@. In section~III the
decay $H^\pm\to H^{\pm\pm}W^*$ is introduced. Section~IV contains our numerical analysis of the magnitude of the cross section for
$H^{++}H^{--}$ which originates from production via $\qqHpmpmHmp$
followed by the decay $H^\pm\to H^{\pm\pm}W^*$.
Conclusions are given in section~V\@.

\section{The Higgs Triplet Model}

In the HTM~\cite{Konetschny:1977bn,Schechter:1980gr,Cheng:1980qt}
a $Y=2$ complex $SU(2)_L$ isospin triplet of 
scalar fields is added to the SM Lagrangian. 
Such a model can provide Majorana masses for the observed neutrinos 
without the introduction of $SU(2)_L$ singlet neutrinos via
the gauge invariant Yukawa interaction:
\begin{equation}
{\cal L}=\hll L_\ell^TCi\tau_2\Delta L_{\ell^\prime}+\text{h.c.}
\label{trip_yuk}
\end{equation}
Here $\hll (\ell,\ell^\prime=e,\mu,\tau)$ is a complex
and symmetric coupling,
$C$ is the Dirac charge conjugation operator,
$\tau_i$ is the Pauli matrix,
$L_\ell=(\nu_{\ell L}, \ell_L)^T$ is a left-handed lepton doublet,
and $\Delta$ is a $2\times 2$ representation of the $Y=2$
complex triplet fields:
\begin{equation}
\Delta
=\bma{cc}
\Delta^+/\sqrt{2}  & \Delta^{++} \\
\Delta^0       & -\Delta^+/\sqrt{2}
\ema .
\end{equation}
A non-zero triplet vacuum expectation value $\langle\Delta^0\rangle$ 
gives rise to the following mass matrix for neutrinos:
\begin{equation}
\mll = 2\hll \langle\Delta^0\rangle = \sqrt{2}\hll v_{\Delta} .
\label{nu_mass}
\end{equation}
The necessary non-zero $v_{\Delta}$ arises from the minimisation of
the most general $SU(2)_L\otimes U(1)_Y$ invariant Higgs potential~\cite{Cheng:1980qt,Gelmini:1980re},
which is written%
\footnote{
 One may rewrite the potential in eq.~(\ref{higgs_potential})
by using
$2 \text{Det}(\Delta^\dagger \Delta)
= [\text{Tr}(\Delta^\dagger \Delta)]^2
- \text{Tr}[(\Delta^\dagger \Delta)^2]$
and
$(\Phi^\dagger \tau_i \Phi) \text{Tr}(\Delta^\dagger \tau_i \Delta)
= 2 \Phi^\dagger \Delta \Delta^\dagger \Phi
-(\Phi^\dagger \Phi) \text{Tr}(\Delta^\dagger \Delta)$.
}
as follows~\cite{Ma:2000wp, Chun:2003ej}
(with $\Phi=(\phi^+,\phi^0)^T$):
\begin{eqnarray}
V&=&m^2(\Phi^\dagger\Phi)+\lambda_1(\Phi^\dagger\Phi)^2+M^2
{\rm Tr}(\Delta^\dagger\Delta) +
\lambda_2[{\rm Tr}(\Delta^\dagger\Delta)]^2+ \lambda_3{\rm Det}
(\Delta^\dagger\Delta)  \nonumber \\
&&+\lambda_4(\Phi^\dagger\Phi){\rm Tr}(\Delta^\dagger\Delta)
+\lambda_5(\Phi^\dagger\tau_i\Phi){\rm Tr}(\Delta^\dagger\tau_i
\Delta)+\left(
{1\over \sqrt 2}\mu(\Phi^Ti\tau_2\Delta^\dagger\Phi) + \text{h.c.} \right) .
\label{higgs_potential}
\end{eqnarray}
Here $m^2<0$ in order to ensure $\langle\phi^0\rangle=v/\sqrt 2$ which
spontaneously breaks $SU(2)\otimes U(1)_Y$
to  $U(1)_Q$, and $M^2\,(>0)$ is the mass term for the triplet scalars.
In the model of Gelmini-Roncadelli~\cite{Gelmini:1980re} 
the term $\mu(\Phi^Ti\tau_2\Delta^\dagger\Phi)$ is absent,
which leads to spontaneous violation of lepton number for $M^2<0$.
The resulting Higgs
spectrum contains a massless triplet scalar (majoron, $J$) and another light 
scalar ($H^0$). Pair production via $e^+e^-\to H^0J$ would give a large 
contribution to the invisible width of the $Z$ and this model
was excluded at the CERN Large Electron Positron Collider (LEP). 
The inclusion of the term $\mu(\Phi^Ti\tau_2\Delta^\dagger\Phi$)~\cite{Cheng:1980qt}
explicitly breaks lepton number $L\#$ when $\Delta$ is assigned $L\#=-2$, and eliminates
the majoron.
Thus the scalar potential in
eq.~(\ref{higgs_potential}) together with the triplet Yukawa interaction of
eq.~(\ref{trip_yuk}) lead to a phenomenologically viable model of neutrino mass
generation.
 For small $v_\Delta/v$,
the expression for $v_\Delta$
resulting from the minimisation of $V$ is:
\begin{equation}
v_\Delta \simeq \frac{\mu v^2}{2M^2+(\lambda_4+\lambda_5)v^2} \ .
\label{tripletvev}
\end{equation}
For large $M$ compared to $v$
one has $v_\Delta \simeq \mu v^2/2M^2$,
which is sometimes referred to as the ``Type II seesaw mechanism''
and would naturally lead to a small $v_\Delta$.
Recently there has been much interest in 
the scenario of light triplet scalars ($M\approx v$) 
within the discovery reach of
the LHC, for which eq.~(\ref{tripletvev}) leads to $v_\Delta\approx \mu$.
In extensions of the HTM the term $\mu(\Phi^Ti\tau_2\Delta^\dagger\Phi$) 
may arise in various ways: i) it can be generated at tree level via the vev of a Higgs singlet field~\cite{Schechter:1981cv}; 
ii) it can arise at higher orders in perturbation theory~\cite{Chun:2003ej};
iii) it can originate in the context of extra dimensions~\cite{Ma:2000wp}.

An upper limit on $v_\Delta$ can be obtained from
considering its effect on the parameter $\rho (=M^2_W/M_Z^2\cos^2\theta_W)$. 
In the SM $\rho=1$ at tree-level, while in the HTM one has
(where $x=v_\Delta/v$):
\begin{equation}
\rho\equiv 1+\delta\rho={1+2x^2\over 1+4x^2} .
\label{deltarho}
\end{equation}
The measurement $\rho\approx 1$ leads to the bound
$v_\Delta/v\lsim 0.03$, or  $v_\Delta\lsim 8\,\GeV$.
Production mechanisms which depend on $v_\Delta$ 
(i.e.\ $pp\to W^{\pm *}\to W^\mp H^{\pm\pm}$ and fusion via 
$W^{\pm *} W^{\pm *} \to H^{\pm\pm}$~\cite{Huitu:1996su,Gunion:1989ci,Vega:1989tt})
are not competitive with the processes 
$\qqHppHmm$ and $\qqHpmpmHmp$
at the energies of the Fermilab Tevatron, but such mechanisms can be the dominant source of 
$H^{\pm\pm}$ at the LHC if $v_{\Delta}={\cal O}(1)\,\GeV$ and 
$m_{H^{\pm\pm}}> 500\,\GeV$.
At the 1-loop level, $v_\Delta$ must be renormalised and explicit
analyses lead to bounds on its magnitude similar to the above bound from
the tree-level analysis, e.g.\ see \cite{Blank:1997qa}.

The scalar eigenstates in the HTM are as follows: i) the charged scalars
 $H^{\pm\pm}$ and $H^\pm$; ii) the CP-even neutral scalars
$h^0$ and $H^0$; iii) a CP-odd neutral scalar $A^0$.
The doubly charged $H^{\pm\pm}$ is entirely composed of the triplet 
scalar field $\Delta^{\pm\pm}$, 
while the remaining eigenstates are in general mixtures of the  
doublet and triplet fields. However, such mixing is proportional to the 
triplet vev, and hence small {\it even if} $v_\Delta$
assumes its largest value of a few GeV\@.%
\footnote{A large mixing angle is possible in the CP-even sector
provided that $m_{h^0}\sim m_{H^0}$~\cite{Akeroyd:2010je,Dey:2008jm}.}
Therefore $H^\pm,H^0,A^0$ are predominantly composed 
of the triplet fields, while $h^0$ is predominantly composed of the 
doublet field and plays the role of the SM Higgs boson.
 The scale of squared masses of $H^{\pm\pm},H^\pm,H^0,A^0$
are determined by $M^2 + \lambda_4 v^2/2$ with mass splittings 
of order $\lambda_5 v^2$~\cite{Ma:2000wp,Chun:2003ej,Akeroyd:2010je}: 
\begin{eqnarray}
m^2_{H^{\pm\pm}}
&\simeq&
 m^2_{H^{\pm}} - \frac{\lambda_5}{2} v^2 ,
\\ \nonumber
m^2_{H^{\pm}}
&\simeq&
 M^2 + \frac{\lambda_4}{2} v^2 ,
\\ \nonumber
m^2_{H^0,A^0}
&\simeq&
 m^2_{H^{\pm}} + \frac{\lambda_5}{2} v^2 .
\label{eq:mH}
\end{eqnarray}
 The degeneracy $m_{H^0} \simeq m_{A^0}$
can be understood by the fact that
the Higgs potential is invariant under a global $U(1)$ for $\Delta$
($L\#$ conservation)
when one neglects the trilinear term proportional to $\mu$.

The mass hierarchy $m_{H^{\pm\pm}} < m_{H^\pm} < 
m_{H^0,A^0}$ is obtained for $\lambda_5 > 0$, and the opposite hierarchy
$m_{H^{\pm\pm}} > m_{H^\pm} > m_{H^0,A^0}$ is obtained for $\lambda_5 < 0$.
In general, one would not expect
degenerate masses for $H^{\pm\pm},H^\pm,H^0,A^0$, but instead one of the above two mass
hierarchies.
 The sign of $\lambda_5$ is not fixed
by theoretical requirements of vacuum stability of the scalar potential~\cite{HTM-potential},
although $|\lambda_5|< 2 m_{H^\pm}^2/v^2$ is necessary
to ensure that $m_{H^{\pm\pm}}^2$ and $m_{H^0,A^0}^2$ in eq.~(\ref{eq:mH}) are positive.
Therefore the decays channels
$H^\pm\to H^{\pm\pm}W^{*}$ and $H^{\pm\pm}\to H^\pm W^{*}$
are possible in the HTM\@.

\section{The decay $H^\pm\to H^{\pm\pm}W^*$ and production of $H^{++}H^{--}$ from 
$\qqHpmpmHmp$
}

The potential importance of the decay channel 
$H^{\pm}\to H^{\pm\pm}W^*$ (for $m_{H^\pm}>m_{H^{\pm\pm}}$) 
has not been quantified in the HTM\@.
For this decay to be kinematically open \cite{Chun:2003ej}
one needs the mass hierarchy where $H^{\pm\pm}$ is the lightest of the triplet scalars
($m_{H^{\pm\pm}} < m_{H^\pm} < m_{H^0,A^0}$), which is obtained for $\lambda_5 > 0$.
For the opposite mass hierarchy with $\lambda_5 < 0$ ($m_{H^{\pm\pm}} > m_{H^\pm} > m_{H^0,A^0}$)
the related decay $H^{\pm\pm}\to H^\pm W^*$ 
was shown to be important in the HTM in \cite{Chakrabarti:1998qy,Chun:2003ej,Akeroyd:2005gt,Perez:2008ha}.
The expression for the decay width of $H^\pm\to H^{\pm\pm}W^*$ is easily obtained from the
expression for  $H^{\pm\pm}\to H^\pm W^*$ by merely interchanging $m_{H^{\pm\pm}}$ and $m_{H^\pm}$.
After summing over all fermion states for  $W^*\to f^\prime \overline{f}$,
excluding the $t$ quark, the decay rate
is given by
\begin{equation}
\Gamma(H^{\pm}\to H^{\pm\pm} W^*\to  H^{\pm\pm} f'\overline f)
\simeq
 \frac{ 9 G_F^2 m_W^4 m_{H^{\pm}} }{ 4\pi^3 }
 \int_{0}^{1-\kappa_{H^{\pm\pm}}^{}}\!\!\!dx_2
 \int_{1-x_2-\kappa_{H^{\pm\pm}}^{}}^{1-\frac{\kappa_{H^{\pm\pm}}^{}}{1-x_2}}
 \!\!\!dx_1 F_{H^{\pm\pm}W}(x_1,x_2) ,
\label{HHWdecay}
\end{equation}
where $\kappa_{H^{\pm\pm}}^{} \equiv m_{H^{\pm\pm}}/ m_{H^\pm}$
and the analytical expression for $F_{ij}(x_1,x_2)$
can be found in \cite{Djouadi:1995gv} (see also \cite{Moretti:1994ds}).
 Note that this decay mode does not depend on $v_\Delta$. In eq.~(\ref{HHWdecay}) we take $f'$ and $\overline f$ to be massless, which is
a good approximation as long as the mass splitting between $m_{H^{\pm\pm}}$ and $m_{H^\pm}$ 
is above the mass of the charmed hadrons ($\sim 2\,\GeV$). In our numerical analysis we will be mostly concerned 
with sizeable mass splittings, $m_{H^\pm}-m_{H^{\pm\pm}}\gg 2\,\GeV$.

The other possible decays for $H^\pm$ are $\Hpmlpmnu$,
$H^\pm\to W^\pm Z$, $H^\pm \to W^\pm h^0$ 
(where $h^0$ is the SM-like scalar field) and $H^\pm \to \overline{t}b$.
Explicit expressions for the decay widths of these channels
can be found in the literature (e.g.\ \cite{Gunion:1989ci, Godbole:1994np, Perez:2008ha}) and they are presented below.
The decay width for $\Hpmlpmnu$ is given by
\begin{eqnarray}
\Gamma(H^\pm \to \ell^\pm \nu)
\equiv
 \sum_{\ell, \ell^\prime}
 \Gamma(\Hpmlpmnu)
\simeq
 \frac{ m_{H^\pm} \sum_i m_i^2 }{ 16 \pi v_\Delta^2 } .
\end{eqnarray}
 Note that $\Gamma(H^\pm \to \ell^\pm \nu)$ has no dependence on 
the neutrino mixing angles because 
$\sum_{\ell, \ell^\prime}|h_{\ell\ell^\prime}|^2 = \sum_i m_i^2/(2v_\Delta^2)$,
where $m_i$~($i=1\text{-}3$) are neutrino masses.
 The decay widths for the channels which are proportional to $v_\Delta^2$ are expressed
as follows:
\begin{eqnarray}
\Gamma(H^\pm \to W^\pm Z)
\simeq
 \frac{ v_\Delta^2 G_F^2 m_{H^\pm}^3 }{ 4 \pi }
 [ \beta(m_{H^\pm}, m_W, m_Z ) ]^3 ,
\label{eq:Gam_Hp_WZ}
\end{eqnarray}
\begin{eqnarray}
\Gamma(H^\pm \to W^\pm h^0)
&\simeq&
 \frac{v_\Delta^2 G_F^2 m_{H^\pm}^3}{4\pi}
 \left(
  \frac{ 2m_{H^\pm}^2 - \lambda_4 v^2 }
       { m_H^2 -m_h^2 }
  - 1
 \right)^2
 [ \beta(m_{H^\pm}, m_W, m_h ) ]^3 ,
\label{eq:Gam_Hp_Wh}
\end{eqnarray}
\begin{eqnarray}
\Gamma(H^\pm \to \overline{t} b)
\simeq
 \frac{3 v_\Delta^2 G_F^2 m_t^2 m_{H^\pm}}{ 2 \pi }
 \left( 1 - \frac{ m_t^2 }{ m_{H^\pm}^2 } \right)^2 ,
\label{eq:Gam_Hp_tb}
\end{eqnarray}
\begin{eqnarray}
\beta(m_1, m_2, m_3)
\equiv
 \sqrt{1-\frac{(m_2+m_3)^2}{m_1^2}}
 \sqrt{1-\frac{(m_2-m_3)^2}{m_1^2}} .
\end{eqnarray}

The decay $H^\pm \to W^\pm h^0$ is caused by
two small mixings of scalar fields.
One is the mixing angle $\theta_\pm \simeq \sqrt{2} v_\Delta/v$
between $\phi^\pm$ and $\Delta^\pm$,
and the other is the mixing angle
$\theta_0 \simeq (2m_{H^\pm}^2 - \lambda_4 v^2) v_\Delta /( (m_{H^0}^2 -m_h^2) v)$
between $\text{Re}(\phi^0)$ and $\text{Re}(\Delta^0)$.
 If $M \gg v$, then 
one has $(2m_{H^\pm}^2 - \lambda_4 v^2)/(m_{H^0}^2 -m_h^2) \simeq 2$
in eq.~(\ref{eq:Gam_Hp_Wh}). Since we are interested in the case
where the exotic scalars have masses of the electroweak scale,
we do not take a very large $M$. However,
we assume $(2m_{H^\pm}^2 - \lambda_4 v^2)/(m_{H^0}^2 -m_h^2) \simeq 2$
for simplicity, which can be achieved by
$\lambda_4 \simeq 2 (m_{H^\pm}^2 - m_{H^0}^2 + m_h^2)/v^2$.
 The decay $H^\pm \to \overline{t} b$ is mediated by
the small $\phi^\pm$ component of $H^\pm$ through $\theta_\pm$.
 For $m_{H^\pm} = {\mathcal O}(100)\,\GeV$,
$\Gamma(H^\pm \to \overline{t} b) \propto m_t^2 m_{H^\pm}$ is comparable to
$\Gamma(H^\pm \to W^\pm Z)$ and $\Gamma(H^\pm \to W^\pm h^0) \propto m_{H^\pm}^3$.
 These three decay widths in eq.~(\ref{eq:Gam_Hp_WZ})-(\ref{eq:Gam_Hp_tb})
are greater than $\Gamma(H^\pm \to \ell \nu)$
for $v_\Delta \gtrsim 0.1\,\MeV$
while $\Gamma(H^\pm \to \ell \nu)$ dominates
for $v_\Delta \lesssim 0.1\,\MeV$.

It has already been shown that the decay $H^{\pm\pm}\to H^{\pm}W^*$ can be the dominant decay channel 
for the doubly charged scalar over a wide range of 
values of $\Delta M \equiv m_{H^{\pm\pm}}-m_{H^\pm}$ and $v_\Delta$%
~\cite{Chakrabarti:1998qy,Chun:2003ej,Akeroyd:2005gt,Perez:2008ha},
even for $\Delta M \ll m_W$.
 Hence we expect a similar result for the decay $H^\pm\to H^{\pm\pm}W^*$ for the singly charged scalar.
 The branching ratio $\BR(H^\pm\to H^{\pm\pm}W^*)$
will be maximised with respect to $v_\Delta$
if
$\Gamma(H^\pm \to \ell^\pm \nu)
 = \Gamma(H^\pm \to W^\pm Z)
 + \Gamma(H^\pm \to W^\pm h^0)
 + \Gamma(H^\pm \to \overline{t} b)$
which is achieved for $v_\Delta \simeq 0.1\,\MeV$.
A numerical study of the magnitude of $\BR(H^\pm\to H^{\pm\pm}W^*)$
is presented in the next section.

We now emphasise an important phenomenological difference between the 
distinct scenarios of a sizeable branching ratio for the decay channels 
 $H^{\pm\pm}\to H^{\pm}W^*$ (for $\lambda_5<0$) and $H^\pm\to H^{\pm\pm}W^*$ (for $\lambda_5 > 0$).
The decay $H^{\pm\pm}\to H^{\pm}W^*$ is expected to
weaken the discovery potential of $H^{\pm\pm}$ at the LHC, because it would reduce the branching ratio of
a channel like $H^{\pm\pm}\to \ell^\pm {\ell^\prime}^\pm$ (which is otherwise the dominant
 channel for $v_\Delta \lsim 0.1\,\MeV$, and enjoys low
SM backgrounds).
We note that there has been no simulation of the detection prospects of $H^{\pm\pm}\to H^{\pm}W^*$, and its signature would be
different to that of the standard decay channels
$H^{\pm\pm}\to \ell^\pm{\ell^\prime}^\pm$ and $H^{\pm\pm}\to W^\pm W^\pm$.

\begin{figure}[t]
\begin{center}
\includegraphics[origin=c, angle=-90, scale=0.5]{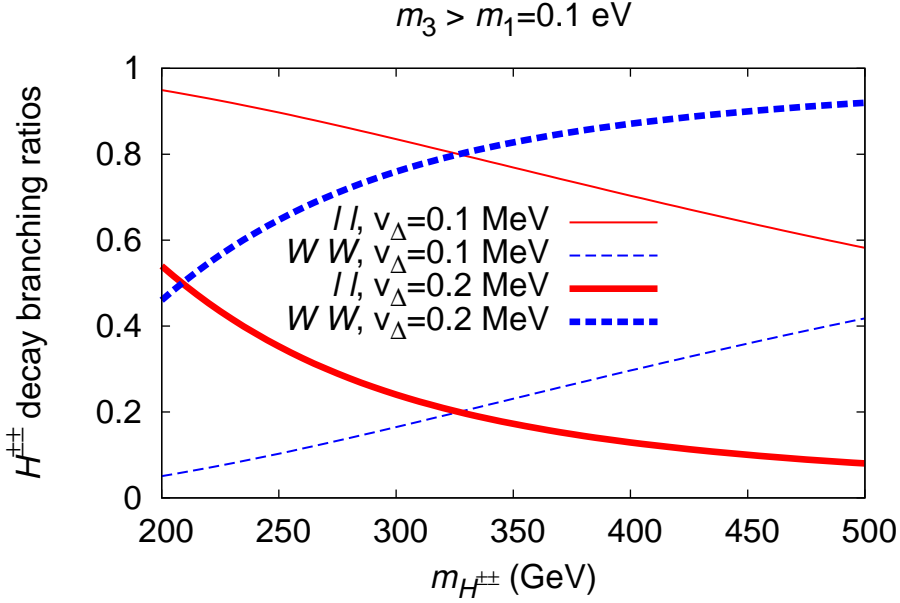}
\vspace*{-25mm}
\caption{
$\sum_{\ell,\ell^\prime} \BR(H^{\pm\pm}\to \ell^\pm{\ell^\prime}^\pm)$ with red solid lines
and 
BR($H^{\pm\pm}\to W^\pm W^\pm)$ with blue dashed lines
as a function of $m_{H^{\pm\pm}}$ ($\leq m_{H^\pm}$)
for $v_\Delta=0.1\,\MeV$ (thin lines)
and $v_\Delta=0.2\,\MeV$ (bold lines).
For the neutrino masses we used $m_1 = 0.1\,\eV$ with $\Delta m^2_{31} > 0$.
}
\label{fig:br_Hdoub}
\end{center}
\end{figure}

In contrast, we point out that the decay $H^{\pm}\to  H^{\pm\pm}W^*$ could actually {\it improve} the discovery potential of
 $H^{\pm\pm}$ at the LHC\@.
From the production mechanism $\qqHpmpmHmp$
the decay mode $H^{\pm}\to  H^{\pm\pm}W^*$
would give rise to pair production ($H^{++}H^{--}$) of doubly charged Higgs bosons.
We believe that this additional way to produce $H^{\pm\pm}$ has not been discussed before.  
In this scenario $H^{\pm\pm}$ is the lightest of the triplet scalars, and its only possible decay channels are
$H^{\pm\pm}\to \ell^\pm{\ell^\prime}^\pm$ and $H^{\pm\pm}\to W^\pm W^\pm$, with branching ratios
determined by the magnitude of $v_\Delta$.
These two branching ratios can be of the same order of magnitude
for $v_\Delta \simeq 0.1\,\MeV$,
as can be seen in Fig.~\ref{fig:br_Hdoub}
where we fix $v_\Delta=0.1\,\MeV$ and $0.2\,\MeV$
(similar figures can be found in \cite{Perez:2008ha}).
 In the range of $m_{H^{\pm\pm}} = 200\to500\,\GeV$,
one has 
$\sum_{\ell,\ell^\prime}\BR(H^{\pm\pm}\to \ell^\pm{\ell^\prime}^\pm) \simeq 100\%$
for $v_\Delta \lesssim 0.05\,\MeV$,
while for $v_\Delta \gtrsim 0.4\,\MeV$
one has BR($H^{\pm\pm}\to W^\pm W^\pm) \simeq 100\%$.

In simulations of pair production of $H^{\pm\pm}$ it is assumed that the production channel 
$\qqHppHmm$ is the only mechanism. If $v_\Delta \lsim 0.1\,\MeV$ then the decay channel
$H^{\pm\pm}\to \ell^\pm {\ell^\prime}^\pm$ is dominant, and four-lepton signatures ($4\ell$)
would be possible. 
Studies have shown that the Standard Model background
for the $4\ell$ signature~\cite{Azuelos:2005uc} is considerably smaller than
that for the signature of $3\ell$~\cite{delAguila:2008cj,Akeroyd:2010ip}, 
and at present it is assumed that the $4\ell$ signature can only arise from
$\qqHppHmm$.
The importance of the production mechanism $\qqHpmpmHmp$ has
been appreciated for the $3\ell$ signature, in which the decay
$H^\mp \to \ell^\mp \nu$ is assumed~\cite{Akeroyd:2005gt,Perez:2008ha,delAguila:2008cj,Akeroyd:2009hb,Akeroyd:2010ip}.
For the case of a sizeable branching ratio for $H^{\pm}\to  H^{\pm\pm}W^*$ we point out 
that the production mechanism $\qqHpmpmHmp$ can also contribute to the $4\ell$ signature, which is the
signature with lowest background. Searches for four leptons originating from $H^{++}H^{--}$
have already been performed by the
Tevatron~\cite{Aaltonen:2008ip} and LHC~\cite{CMS-search}.
If BR($H^{\pm}\to  H^{\pm\pm}W^*$) were sizeable we would expect a 
strengthening of the derived limit on $m_{H^{\pm\pm}}$.

\section{Numerical Analysis} 
In this section we quantify the magnitude of the number of pair-produced $H^{++}H^{--}$ arising from
the process $\qqHpmpmHmp$ with decay $H^{\pm}\to  H^{\pm\pm}W^*$, and make a comparison
with the conventional mechanism $\qqHppHmm$.

 The important parameters for our analyses are
$v_\Delta$, $m_{H^\pm}$, and $m_{H^{\pm\pm}}$.
 We take $m_{H^{\pm\pm}}=200\,\GeV$ or $500\,\GeV$
and show results as functions of $v_\Delta$ and $m_{H^\pm}$.
 The decay branching ratios of $H^\pm$
also depend on two undetermined parameters,
$m_h$ and $m_1$ (one of the neutrino masses).
These are fixed as $m_h=120\,\GeV$ and $m_1 = 0.1\,\eV$ in our numerical analysis.
 Note that $m_h$ only enters through the
decay width for $H^\pm\to W^\pm h^0$.
 Neutrino oscillation experiments%
~\cite{solar,atm,acc,Apollonio:2002gd,:2008ee} provide a measurement of two
neutrino mass differences,
$\Delta m^2_{ij} \equiv m_i^2 - m_j^2$,
and we use the following values:
$\Delta m^2_{21} = 7.6\times 10^{-5}\,\eV$ ,
$|\Delta m^2_{31}| = 2.4\times 10^{-3}\,\eV$.
 Although $\Delta m^2_{31} > 0$ (referred to as ``normal mass ordering'') is also assumed in our analysis,
our results do not change significantly for $\Delta m^2_{31} < 0$
because the neutrino masses are almost degenerate for $m_1=0.1\,\eV$.

 The experimental bound $\BR(\meee) < 1.0\times 10^{-12}$
gives a stringent constraint on 
 $|\hll|$ and $m_{H^{\pm\pm}}$.\footnote{
 This stringent constraint can be avoided
if $|m_{e\mu}|=0$~\cite{Chun:2003ej}
or $|m_{ee}|=0$~\cite{Akeroyd:2009nu}.
 See also \cite{Fukuyama:2009xk}.}
 Assuming naively
$|m_{ee}| \simeq |m_{e\mu}| \simeq m_1$ for $m_1 = 0.1\,\eV$,
the bound on
$\BR(\meee) = |m_{ee}|^2 |m_{e\mu}|^2/ (16 G_F^2 v_\Delta^4 m_{H^{\pm\pm}}^4)$
can be translated into the constraint
$v_\Delta m_{H^{\pm\pm}} \gtrsim 1.5\times 10^4\,\eV\cdot\GeV$.
 Therefore,
we use $v_\Delta \geq 100\,\eV$
in order to satisfy this constraint for $m_{H^{\pm\pm}}=200\,\GeV$.

\begin{figure}[t]
\begin{center}
\includegraphics[origin=c, angle=-90, scale=0.32]{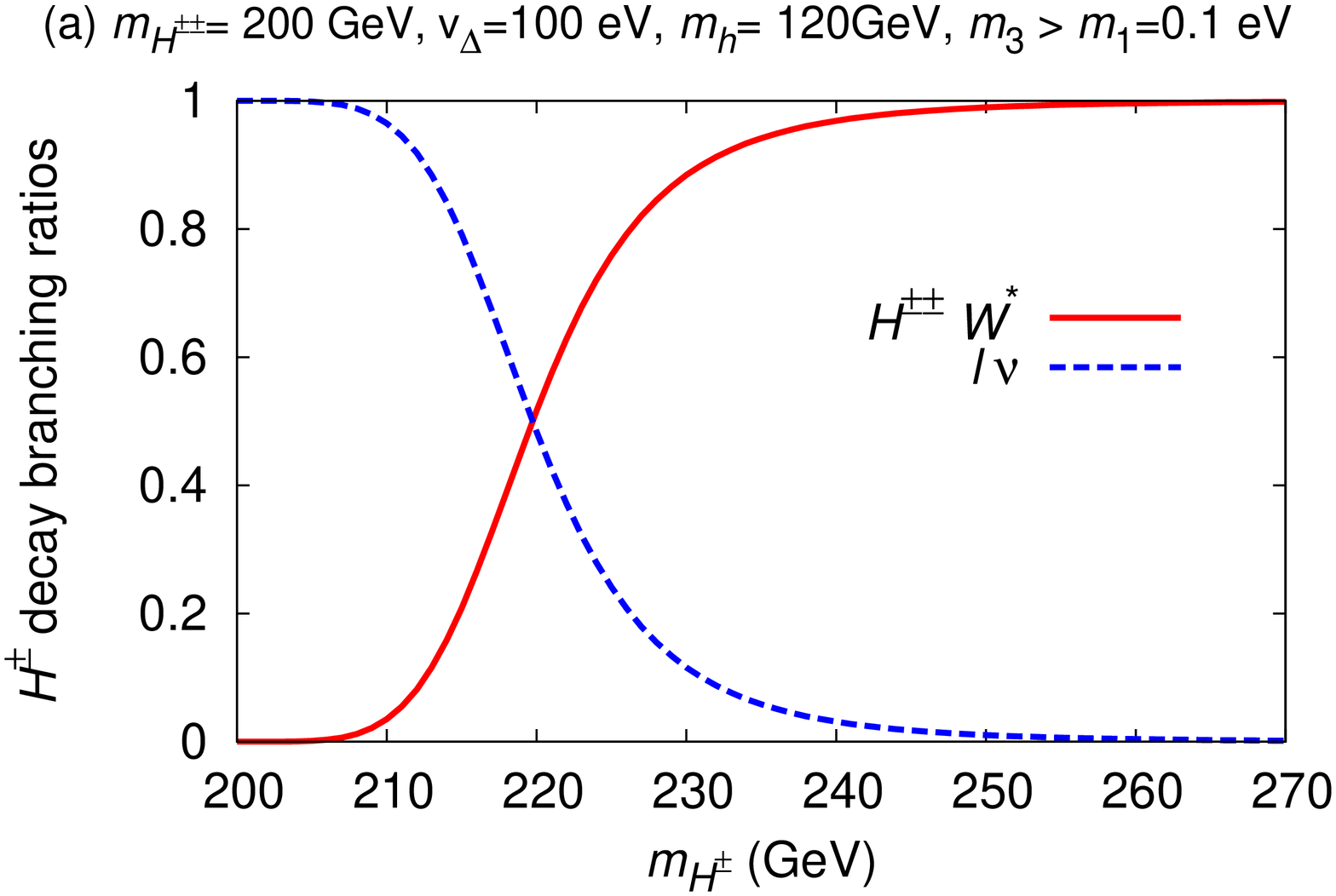}
\includegraphics[origin=c, angle=-90, scale=0.32]{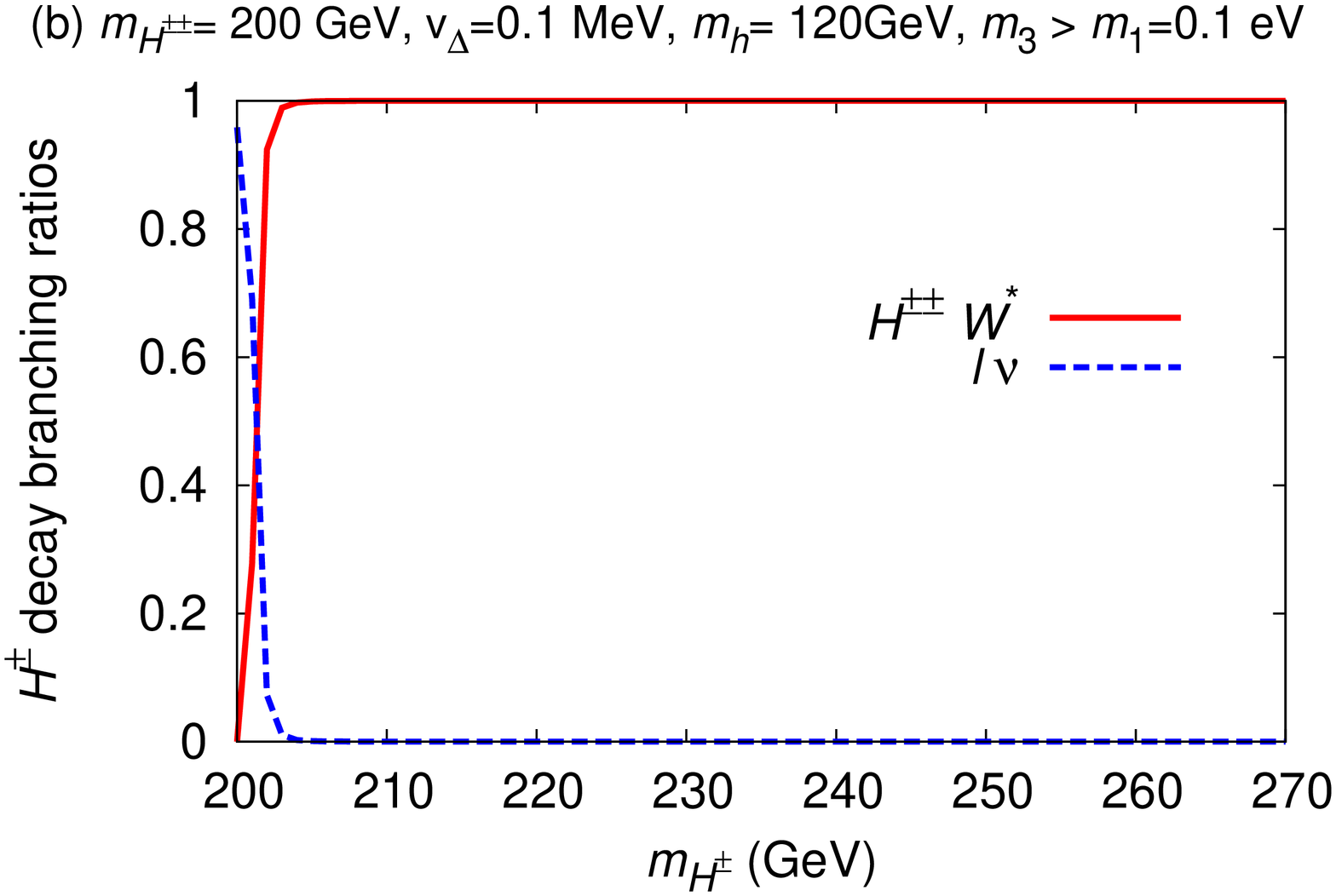}\\[-10mm]
\includegraphics[origin=c, angle=-90, scale=0.32]{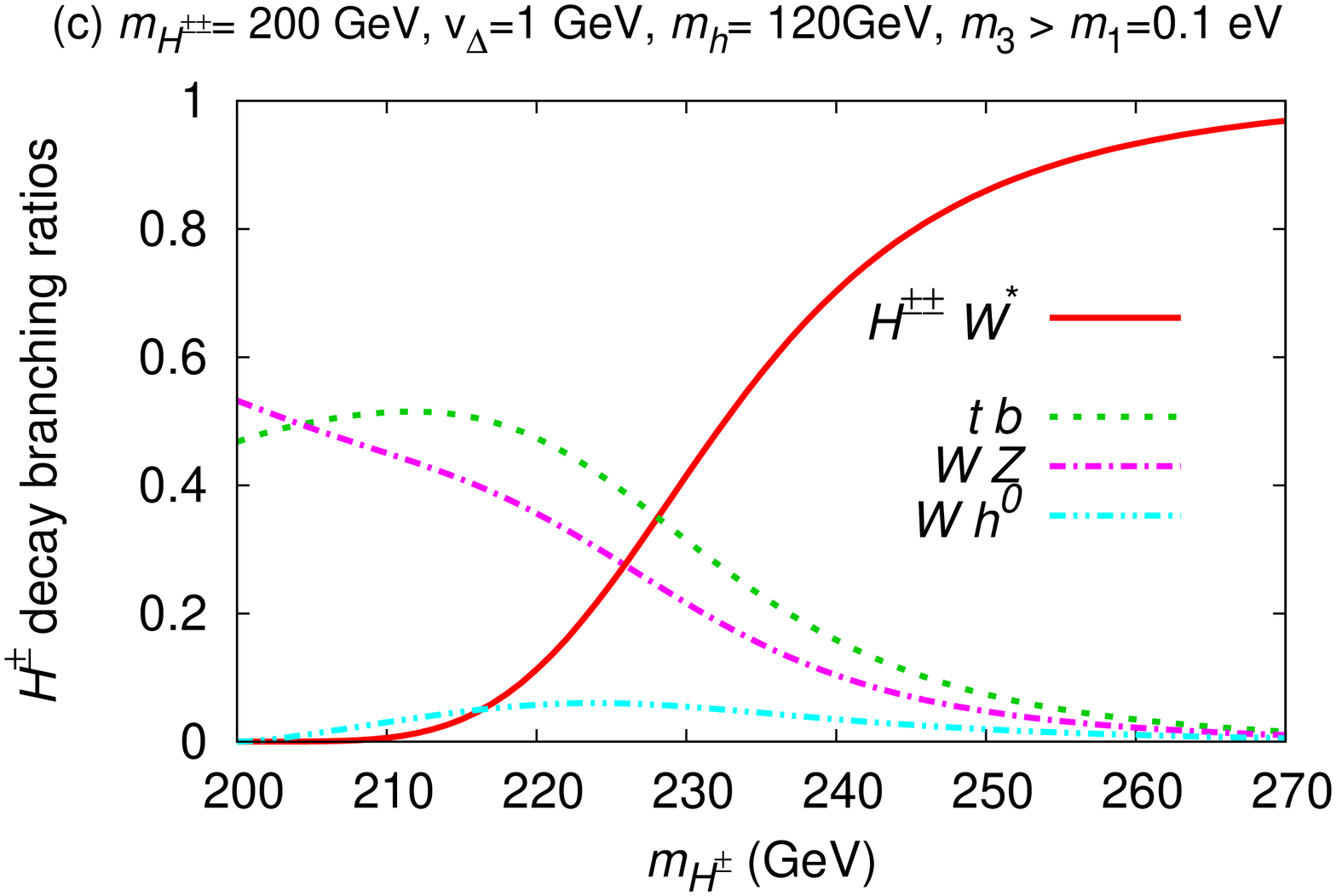}
\vspace*{-15mm}
\caption{
The BRs of $H^\pm$ decays
into $H^{\pm\pm} W^\ast$ (red solid line),
$\ell \nu$ (blue dashed line),
$\bar{t} b$ (green dotted line),
$W Z$ (magenta dot-dashed line),
and $W h^0$ (cyan dot-dot-dashed line)
as a function of $m_{H^{\pm}}$.
In all panels $m_{H^{\pm\pm}}=200\,\GeV$ and $m_{h^0}=120\,\GeV$. In panels a), b) and c)
we fix $v_\Delta=100\,\eV$, $0.1\,\MeV$ and $1\,\GeV$ respectively.
For the neutrino masses we used $m_1 = 0.1\,\eV$ with $\Delta m^2_{31} > 0$.
 In each panel the
decay modes with a negligible BR are omitted.
}
\label{fig:br_Hsing}
\end{center}
\end{figure}

\begin{figure}[t]
\begin{center}
\includegraphics[origin=c, angle=-90, scale=0.5]{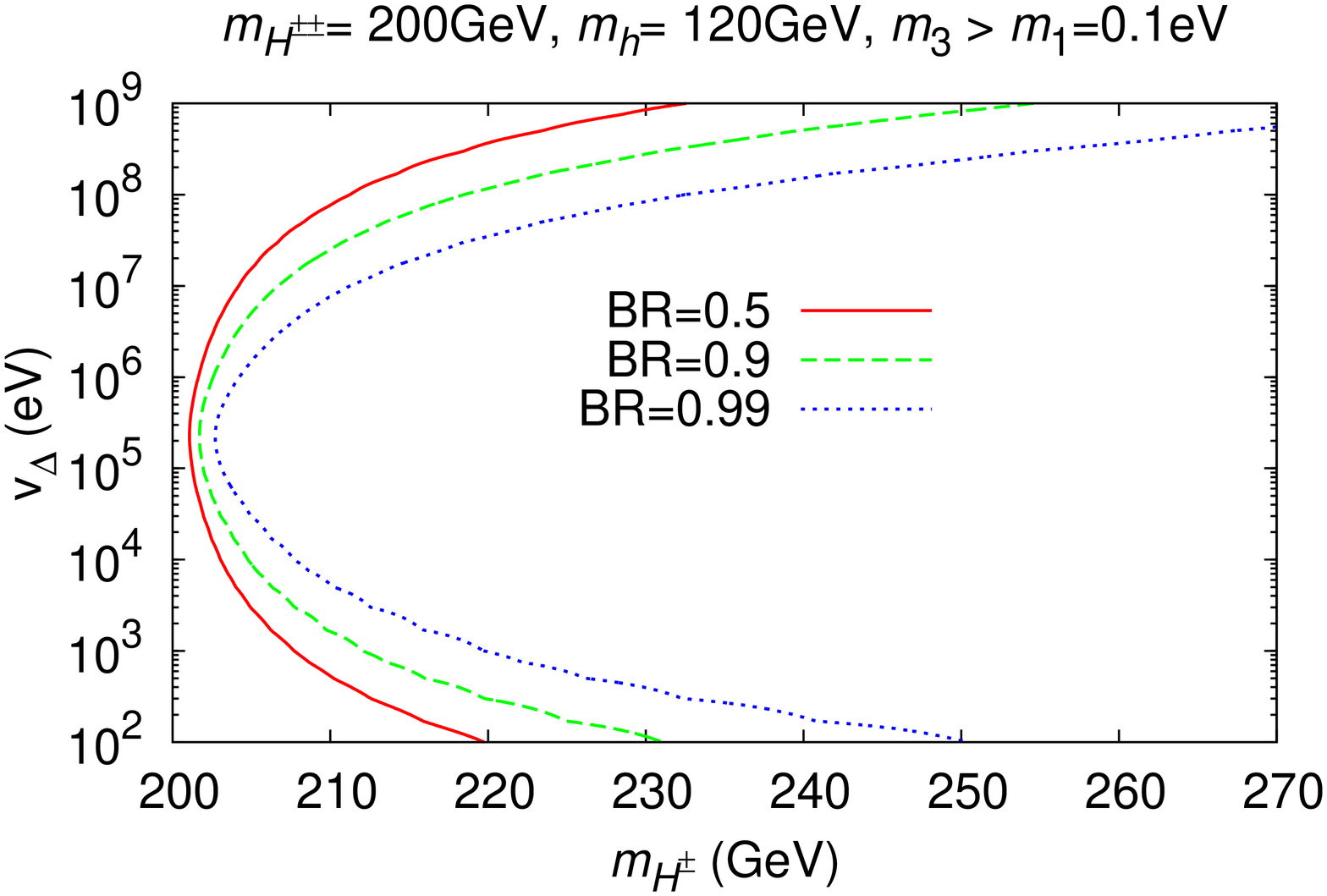}
\vspace*{-25mm}
\caption{
 Contours of $\BR(H^\pm\to H^{\pm\pm} W^\ast)$
in the plane $(m_{H^\pm}, v_\Delta)$.
 We used $m_1 = 0.1\,\eV$ with $\Delta m^2_{31} > 0$.
The red solid, green dashed, and blue dotted lines
show contours of
$\BR(H^\pm\to H^{\pm\pm} W^\ast)=0.5$, 0.9 and 0.99, respectively.
}
\label{fig:br_contour}
\end{center}
\end{figure}
 In Fig.~\ref{fig:br_Hsing}
we show the BRs of $H^\pm$ decays
into $H^{\pm\pm} W^\ast$ (red solid),
$\ell \nu$ (blue dashed),
$\bar{t} b$ (green dotted),
$W Z$ (magenta dot-dashed),
and $W h^0$ (cyan dot-dot-dashed)
as a function of $m_{H^{\pm}}$
for various values of $v_\Delta$,
fixing $m_{H^{\pm\pm}}=200\,\GeV$ and $m_{h^0}=120\,\GeV$.
 The range of $m_{H^\pm}$ in the figures
corresponds to $0\leq \lambda_5 \lesssim 1$,
which easily satisfies the perturbative constraint $\lambda_5< 4\pi$.
Very large mass splittings (e.g.\ $\gg 100\,\GeV$) are constrained by measurements of
electroweak precision observables, but the mass splittings in Fig.~\ref{fig:br_Hsing}
are compatible with the analyses in \cite{Blank:1997qa} (which are for models with a $Y=0$ triplet). 
 In Fig.~\ref{fig:br_Hsing}(a)
we fix $v_\Delta=100\,\eV$,
for which $\sum_{\ell,\ell^\prime} \BR(H^{\pm\pm} \to \ell^\pm {\ell^\prime}^\pm) \simeq 100\,\%$.
 One can see that
$H^{\pm}\to H^{\pm\pm}W^*$ competes with $H^\pm \to \ell^\pm\nu$,
with all other decay channels being negligible.
 For $|\Delta M|>20\,\GeV$,
$H^{\pm}\to H^{\pm\pm}W^*$ becomes the dominant decay channel.
 In Fig.~\ref{fig:br_Hsing}(b)
we fix $v_\Delta=0.1\,\MeV$,
and $H^{\pm}\to H^{\pm\pm}W^*$ becomes the dominant decay channel
for much smaller mass splittings,
$|\Delta M|>2\,\GeV$.
 In Fig.~\ref{fig:br_Hsing}(c)
we fix $v_\Delta=1\,\GeV$, for which
the competing decays are $H^\pm \to tb$, $H^\pm \to WZ$ and $H^\pm \to Wh^0$.
 In this scenario
the decay $H^{\pm}\to H^{\pm\pm}W^*$ becomes the dominant channel for 
$|\Delta M|>30\,\GeV$.

In Fig.~\ref{fig:br_contour}
we show contours of BR($H^\pm\to H^{\pm\pm}W^*$) in the plane
$[m_{H^\pm}, v_\Delta]$.
The red solid, green dashed, and blue dotted lines
correspond to contours of
$\BR(H^\pm\to H^{\pm\pm} W^\ast)=0.5$, 0.9, and 0.99, respectively.
 The BR is maximised
at around $v_\Delta = 0.1\,\MeV$, as expected.
It is clear from Fig.~\ref{fig:br_Hsing} and Fig.~\ref{fig:br_contour} 
that the decay of $H^\pm$ into $H^{\pm\pm}$ can be dominant
in a wide region of the parameter space of the HTM
even if the two-body decay into $H^{\pm\pm} W^\mp$ (for $m_{H^\pm}>m_{H^{\pm\pm}}+m_W$)
is forbidden kinematically.
Moreover, for $v_\Delta = 100\,\eV$ (i.e.\ when the four-lepton signal
arising from the decay of $H^{++}H^{--}$ is dominant) the magnitude of
$\BR(H^\pm\to H^{\pm\pm} W^\ast)$
becomes very large if $|\Delta M| \gtrsim 30\,\GeV$.

\begin{figure}[t]
\begin{center}
\includegraphics[origin=c, angle=-90, scale=0.32]{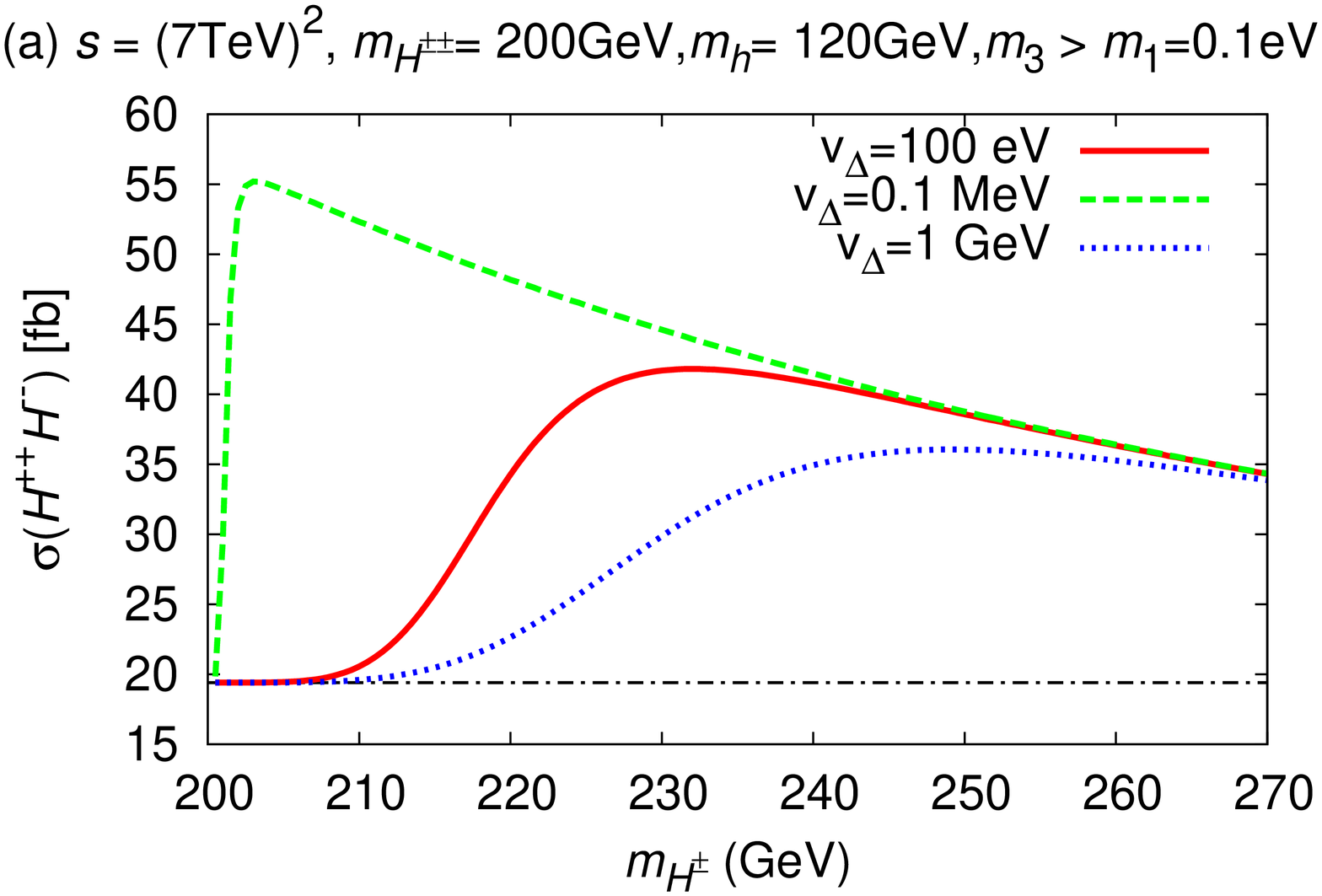}
\includegraphics[origin=c, angle=-90, scale=0.32]{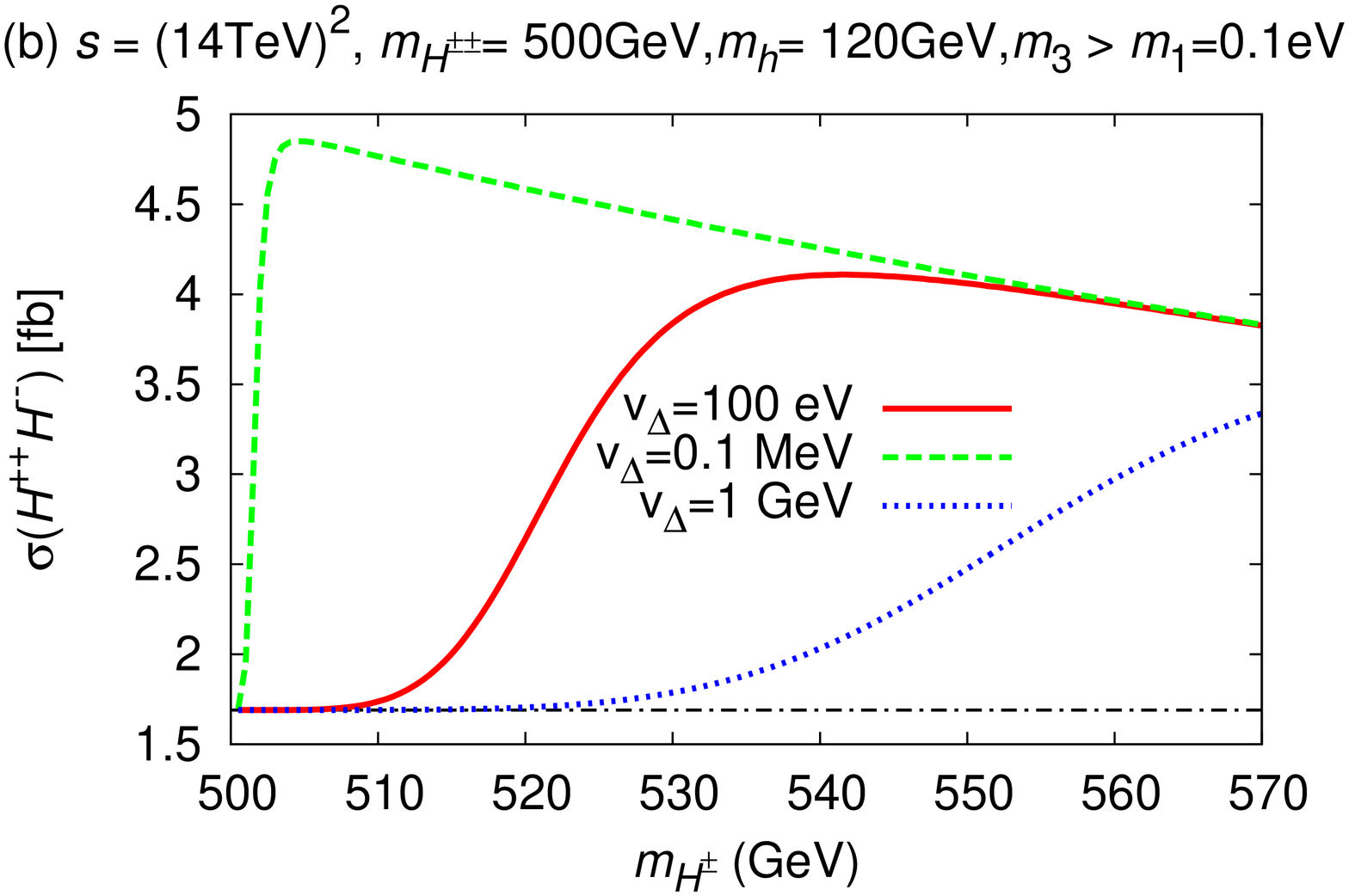}
\vspace*{-20mm}
\caption{
 Behaviour of $\sigma(H^{++}H^{--})\equiv \sigma(\ppHppHmm) + X_1$
as a function of $m_{H^\pm}$.
(a) for $m_{H^{\pm\pm}}=200\,\GeV$ at the LHC with $\sqrt{s} = 7\,\TeV$.
(b) for $m_{H^{\pm\pm}}=500\,\GeV$ at the LHC with $\sqrt{s} = 14\,\TeV$.
 We used $m_h=120\,\GeV$ and $m_1 = 0.1\,\eV$ with $\Delta m^2_{31} > 0$.
The red solid, green dashed, and blue dotted lines
show results for
$v_\Delta=100\,\eV$, $0.1\,\MeV$,
and $1\,\GeV$, respectively.
 The horizontal dot-dashed line shows
$\sigma(\ppHppHmm)$.
}
\label{fig:brsig}
\end{center}
\end{figure}

We now study the magnitude of the number of pair-produced $H^{++}H^{--}$ 
which originate from
$\ppHpmpmHmp$ followed by the decay $H^\pm\to H^{\pm\pm}W^*$.
We define the variable $X_1$ as follows:
\begin{eqnarray}
X_1
\equiv
 \Bigl\{ \sigma(\ppHppHm) + \sigma(\ppHmmHp) \Bigr\}
 BR(H^\pm\to H^{\pm\pm}W^*) .
\label{eq:HppHm}
\end{eqnarray}
 In Fig.~\ref{fig:brsig}
we show the behaviour of
$\sigma(H^{++}H^{--})\equiv \sigma(\ppHppHmm) + X_1$
with respect to $m_{H^\pm}$
for several values of $v_\Delta$. In Fig.~\ref{fig:brsig}a we take $m_{H^{\pm\pm}}=200\,\GeV$
and $\sqrt{s}=7\,\TeV$, and in Fig.~\ref{fig:brsig}b we take $m_{H^{\pm\pm}}=500\,\GeV$
and $\sqrt{s}=14\,\TeV$. We use CTEQ6L1 parton distribution functions \cite{Pumplin:2002vw}.
 The range of $m_{H^\pm}$ in Fig.~\ref{fig:brsig}b
corresponds to $0\leq \lambda_5 \lesssim 2.5$.
 The horizontal dot-dashed line corresponds to the case
of $X_1=0$, i.e.\ the magnitude of $\sigma(\ppHppHmm)$ alone.
 The red solid, green dashed, and blue dotted lines
are the results with $v_\Delta = 100\,\eV$, $0.1\,\MeV$,
and $1\,\GeV$, respectively.
 The red solid line (for which
$\sum_{\ell, \ell^\prime} \BR(H^{\pm\pm}\to \ell^\pm{\ell^\prime}^\pm)\simeq 100\%$)
shows that
the extra contribution from $H^\pm\to H^{\pm\pm}W^*$
can enhance the number of four-lepton events
by a factor of 2 (at $m_{H^\pm}\simeq 230\,\GeV$ in Fig.~\ref{fig:brsig}a)
and 2.4 (at $m_{H^\pm}\simeq 540\,\GeV$ in Fig.~\ref{fig:brsig}b).
For  $v_\Delta = 0.1\,\MeV$, around which
$\sum_{\ell, \ell^\prime} \BR(H^{\pm\pm}\to \ell^\pm{\ell^\prime}^\pm)$
can still be sizeable (See Fig.~\ref{fig:br_Hdoub}),
the enhancement factor for pair-produced $H^{++}H^{--}$ can be 
as large as 2.6 in Fig.~\ref{fig:brsig}a and 2.8 in Fig.~\ref{fig:brsig}b.
 For $v_\Delta=1\,\GeV$ 
the enhancement of pair-produced $H^{++}H^{--}$
is interpreted as an increase in the number of $W^+W^+W^-W^-$ events,
because $\BR(H^{\pm\pm}\to W^\pm W^\pm)\simeq 100\%$.
The shape of the curves is caused by the different
dependence of the cross section and BR on the mass splitting
$\Delta M$.
As $m_{H^\pm}$ increases, the cross section of
$\ppHppHmm$ is unaffected but the cross section
of $\ppHpmpmHmp$ decreases. However,
a larger mass splitting is favourable from the point of view of the BR\@.

Finally, 
we note that a pair of $H^{\pm\pm}$ can also be produced from
other production mechanisms, namely $H^+H^-$, $H^\pm H^0$, $H^\pm A^0$, and $H^0 A^0$.
Although the contribution from $H^{\pm\pm}H^\mp$ in eq.~(\ref{eq:HppHm})
is the most important one
because of the mass hierarchy $m_{H^{\pm\pm}}<m_{H^\pm}<m_{H^0,A^0}$
and its linear dependence on 
BR$(H^\pm\to H^{\pm\pm}W^*)$, the above mechanisms can give
a significant contribution to the number of pair-produced  $H^{\pm\pm}$, as will
be described qualitatively below.

Naively, one would expect the next most important mechanism to be  $H^+H^-$
because its contribution to the production of $H^{++}H^{--}$ scales as $\BR^2$ as
follows:
\begin{eqnarray}
X_2
\equiv
 \sigma(\ppHpHm) [\BR(H^\pm\to H^{\pm\pm} W^*)]^2 .
\end{eqnarray}
 However,
the couplings for $\gamma H^+H^-$ and $Z H^+H^-$
are about a half of those for $\gamma H^{++}H^{--}$ and $Z H^{++}H^{--}$,
respectively.
 The interference between $\gamma^\ast$ and $Z^\ast$ is destructive
for $H^+H^-$ production while it is constructive for $H^{++}H^{--}$ production.
 Tables~\ref{tab:LHC7} and \ref{tab:LHC14} show that
$\sigma(\ppHpHm)$ is smaller than
$\sigma(\ppHppHmm)$ by a factor of $\sim 9.5$, even for $m_{H^\pm}=m_{H^{\pm\pm}}$
(see e.g.\ \cite{Davidson:2010sf}).
Moreover, $X_2$ is suppressed relative to $X_1$ by an extra factor
of BR when $\BR(H^\pm\to H^{\pm\pm} W^*)\neq 100\%$.
Therefore the contribution from $\sigma(\ppHpHm)$ to the production of 
$H^{++}H^{--}$ is considerably less than the QCD $K$ factor for 
$\ppHppHmm$ (which is known to be around 1.25 at the LHC~\cite{Muhlleitner:2003me}).

It turns out that
the production of $H^\pm H^0$ and $H^\pm A^0$ are
numerically more important than $H^+ H^-$,
despite their contributions scaling as $\BR^3$.
 The narrow width approximation for contributions
from $H^0$ and $A^0$ with $m_{H^0} \simeq m_{A^0}$
is rather complicated because of their interference.
We define the variables $X_3$ and $X_3^\prime$ as follows: 
\begin{eqnarray}
X_3
&\equiv&
 \left\{
  \sigma(\ppHpH) + \sigma(\ppHmH)
 \right\}
\nonumber\\
&&\hspace*{50mm}
\times
 \BR_+\,
 [\BR(H^\pm \to H^{\pm\pm} W^*)]^2 ,
\label{eq:HpH0}
\\
X_3^\prime
&\equiv&
 \left\{
  \sigma(\ppHpH) + \sigma(\ppHmH)
 \right\}
\nonumber\\
&&\hspace*{50mm}
\times
 \BR_-\,
 [\BR(H^\pm \to H^{\pm\pm} W^*)]^2 ,
\label{eq:HpH0-2}
\\
%
%
\BR_\pm
&\equiv&
 \BR(H^0\to H^\pm W^*)
 +
 \BR(A^0\to H^\pm W^*)
\nonumber\\
&&\hspace*{20mm}
{}\pm
 \frac{ 4 \BR(H^0\to H^\pm W^*) \BR(A^0\to H^\pm W^*) }
      { \BR(H^0\to H^\pm W^*) + \BR(A^0\to H^\pm W^*) } ,
\end{eqnarray}
where we used
$\sigma(\ppHpmA) \simeq \sigma(\ppHpmH)$
because $m_{A^0} \simeq m_{H^0}$.
The interesting point is that $X_3^\prime$ is for the process
which gives {\it same-sign} $H^{++}H^{++}$ (with $3(W^-)^\ast$)
and $H^{--}H^{--}$ (with $3(W^+)^\ast$)
while $X_3$ is for $H^{++}H^{--}$ production.
 Since $X_3^\prime$ arises as the breaking effect
of the lepton number ($\Delta$ has $L\# = -2$),
it vanishes for $v_\Delta\to 0$,
for which the total decay widths satisfy
$\Gamma_{\text{tot}}(H^0) = \Gamma_{\text{tot}}(A^0)$,
namely $\BR(H^0\to H^\pm W^*) = \BR(A^0\to H^\pm W^*)$.
 This means that the {\it same-sign} $H^{\pm\pm}H^{\pm\pm}$
would not give the {\it same-sign} $4\ell$ signal
because $\BR(H^{\pm\pm} \to \ell^\pm{\ell^\prime}^\pm)$
is small for a large $v_\Delta$
where $X_3^\prime$ could be sizeable.
 A pair of $H^{\pm\pm}$ ({\it same-sign} or different sign)
is provided by $X_3+X_3^\prime$,
which is proportional to
$2[\BR(H^0\to H^\pm W^*)+\BR(A^0\to H^\pm W^*)]$;
 the factor of 2 compensates the fact that
the sum of the cross sections in eq.~(\ref{eq:HpH0})
is a half of the sum in eq.~(\ref{eq:HppHm})
for $m_{H^0,A^0} = m_{H^\pm} = m_{H^{\pm\pm}}$
as shown in Tables~\ref{tab:LHC7} and \ref{tab:LHC14}.
 Although
$\BR(A^0\to H^+ W^*)=\BR(A^0\to H^- W^*)$ (likewise for $H^0$)
and the maximum value of each is $50\%$,
this is compensated by $\BR(H^0\to H^\pm W^*)+\BR(A^0\to H^\pm W^*)$
in $X_3+X_3^\prime$.
Since the partial decay widths of $H^0$ and $A^0$
depend on the scalar masses and $v_\Delta$ in a way which is very similar to 
the partial decay widths of $H^\pm$ (see e.g.\ \cite{Perez:2008ha}),
the analogies of Fig.~\ref{fig:br_contour}
for $\BR(A^0\to H^\pm W^*)$ and $\BR(H^0\to H^\pm W^*)$
would show a similar quantitative behaviour as Fig.~\ref{fig:br_contour}.\footnote{We note that
the decays $A^0\to H^\pm W^*$ and $H^0\to H^\pm W^*$ were also mentioned as a source of
$H^\pm$ in \cite{Chun:2003ej}.}
Thus the main difference between $X_1$ and $X_3+X_3^\prime$
would be the phase space factor because we take $m_{H^0,A^0} > m_{H^\pm}$.
The contribution of $X_3+X_3^\prime$ to the production of a pair of $H^{\pm\pm}$
would be sizeable for $v_\Delta \simeq 0.1\,\MeV$,
where the relevant BRs in eq.~(\ref{eq:HpH0})
could be very large for a small mass splitting.
 Moreover,
the contribution of $X_3+X_3^\prime$ would not be so small
even for large mass splittings
e.g.\ $m_{H^\pm} = 270\,\GeV$ and $m_{H^{\pm\pm}} = 200\,\GeV$
(which give $m_{H^0,A^0} = 325\,\GeV$),
for which the BRs in eq.~(\ref{eq:HpH0}) could be maximal.

 The last mechanisms (which scale as $\BR^4$) are
\begin{eqnarray}
X_4
&\equiv&
 \sigma(\ppHA)\,
 \BR_+^2\,
 [\BR(H^\pm\to H^{\pm\pm} W^*)]^2 ,
\\
X_4^\prime
&\equiv&
 \sigma(\ppHA)\,
 \BR_+\, \BR_-\,
 [\BR(H^\pm\to H^{\pm\pm} W^*)]^2 ,
\\
X_4^{\prime\prime}
&\equiv&
 \sigma(\ppHA)\,
 \BR_-^2\,
 [\BR(H^\pm\to H^{\pm\pm} W^*)]^2 .
\end{eqnarray}
Note that $X_4^\prime$ gives a pair of {\it same-sign} $H^{\pm\pm}$
(being proportional to $\BR_-$, like $X_3^\prime$) and its magnitude is negligible for small $v_\Delta$.
 Although both of $X_4$ and $X_4^{\prime\prime}$ give $H^{++}H^{--}$,
$X_4^{\prime\prime}$ also vanishes for $v_\Delta\to 0$
because it is sensitive to $\BR_-^2$ i.e.\ it is quadratic in lepton number violation.
 The phase space suppression ($m_{H^{\pm\pm}}<m_{H^\pm}<m_{H^0,A^0}$)
ensures that
$\sigma(\ppHA)$ is much smaller than
$\sigma(\ppHppHmm)$
for the case of a large mass splitting with $\sqrt{s}=7\,\TeV$.
Therefore, for $X_4$ to be important a large mass splitting 
with $\sqrt{s}=14\,\TeV$ or a small mass splitting for $v_\Delta \simeq 0.1\,\MeV$
are preferred.

We note that the detection efficiencies for the above mechanisms
($X_1$, $X_2$, $X_3$ and $X_4$) would in general be different from that of the
well-studied mechanism $\ppHppHmm$ because of the extra $W^\ast$.
We defer a detailed study to a future work.

\begin{table}[t]
\begin{tabular}{c||c|c|c|c|c|c|c|}
 $\sqrt{s}=7\,\TeV$
  & \multicolumn{7}{|c|}{ $\sigma(pp \to V^\ast \to H_1 H_2)$\,[fb] }
\\\cline{2-8}
 $m_{H^{\pm\pm}}=200\,\GeV$
  & \ $H^{++}H^{--}$ \
  & \ $H^{++}H^-$ \
  & \ $H^{--}H^+$ \
  & \ $H^+ H^-$ \
  & \ $H^+ H^0$ \
  & \ $H^- H^0$ \
  & \ $H^0 A^0$
\\
 {}
  &
  &
  &
  &
  & \ (or $H^+ A^0$) \
  & \ (or $H^- A^0$) \
  &
\\\hline\hline
 \ $m_{H^\pm} = 200\,\GeV$ \
  & $19$
  & $26$
  & $10$
  & $2.0$
  & $13$
  & $5.1$
  & $18$
\\\hline
 \ $m_{H^\pm} = 230\,\GeV$ \
  & $19$
  & $19$
  & $7.3$
  & $1.1$
  & $5.6$
  & $2.1$
  & $6.0$
\\\hline
 \ $m_{H^\pm} = 270\,\GeV$ \
  & $19$
  & $13$
  & $4.8$
  & $0.54$
  & $2.2$
  & $0.76$
  & $1.9$
\end{tabular}
\caption{
 Production cross sections of a pair of exotic Higgs bosons ($H_1 H_2$)
from off-shell gauge bosons ($V^\ast$) in the HTM
at the LHC with $\sqrt{s}=7\,\TeV$.
 We take $m_{H^{\pm\pm}} = 200\,\GeV$ and
we use a relation $m_{H^0,A^0}^2 = 2 m_{H^\pm}^2 - m_{H^{\pm\pm}}^2$;
$m_{H^0,A^0}=200$, $257$, $325\,\GeV$ for
$m_{H^\pm} = 200$, $230$, $270\,\GeV$, respectively.
}
\label{tab:LHC7}
\end{table}

\begin{table}[t]
\begin{tabular}{c||c|c|c|c|c|c|c|}
 $\sqrt{s}=14\,\TeV$
  & \multicolumn{7}{|c|}{ $\sigma(pp \to V^\ast \to H_1 H_2)$\,[fb] }
\\\cline{2-8}
 $m_{H^{\pm\pm}}=500\,\GeV$
  & \ $H^{++}H^{--}$ \
  & \ $H^{++}H^-$ \
  & \ $H^{--}H^+$ \
  & \ $H^+ H^-$ \
  & \ $H^+ H^0$ \
  & \ $H^- H^0$ \
  & \ $H^0 A^0$
\\
 {}
  &
  &
  &
  &
  & \ (or $H^+ A^0$) \
  & \ (or $H^- A^0$) \
  &
\\\hline\hline
 \ $m_{H^\pm} = 500\,\GeV$ \
  & $1.7$
  & $2.3$
  & $0.83$
  & $0.18$
  & $1.1$
  & $0.42$
  & $1.5$
\\\hline
 \ $m_{H^\pm} = 540\,\GeV$ \
  & $1.7$
  & $1.9$
  & $0.69$
  & $0.13$
  & $0.69$
  & $0.24$
  & $0.78$
\\\hline
 \ $m_{H^\pm} = 570\,\GeV$ \
  & $1.7$
  & $1.7$
  & $0.60$
  & $0.097$
  & $0.49$
  & $0.17$
  & $0.50$
\end{tabular}
\caption{
 Production cross sections of a pair of exotic Higgs bosons ($H_1 H_2$)
from off-shell gauge bosons ($V^\ast$) in the HTM
at the LHC with $\sqrt{s}=14\,\TeV$.
 We take $m_{H^{\pm\pm}} = 500\,\GeV$ and
we use a relation $m_{H^0,A^0}^2 = 2 m_{H^\pm}^2 - m_{H^{\pm\pm}}^2$;
$m_{H^0,A^0}=500$, $577$, $632\,\GeV$ for
$m_{H^\pm} = 500$, $540$, $570\,\GeV$, respectively.
}
\label{tab:LHC14}
\end{table}

\section{Conclusions}
Doubly charged Higgs bosons ($H^{\pm\pm}$), which arise in the Higgs Triplet Model (HTM)
of neutrino mass generation, are being searched for at the Tevatron and at the LHC\@.
We showed that $H^{\pm\pm}$ can be produced from the decay of a singly charged Higgs boson ($H^{\pm}$)
via $H^\pm \to H^{\pm\pm}W^*$, which can have a large branching ratio in a wide region of the
parameter space of the HTM\@. From the production mechanism $q'\overline q\to
W^* \to H^{\pm\pm}H^{\mp}$, the above decay would give rise to pair production $H^{++}H^{--}$, with
a number of events which can be comparable to that from the conventional mechanism
$\qqHppHmm$.
Current simulations and searches for $H^{++}H^{--}$ at the Tevatron/LHC assume
production solely from $\qqHppHmm$.
The contribution from $\qqHpmpmHmp$ with decay $H^\pm \to H^{\pm\pm}W^*$ would be
an additional source of pair-produced $H^{\pm\pm}$, which should enhance the detection prospects
in this channel (e.g.\ four-lepton signatures if the decay mode $H^{\pm\pm}\to \ell^\pm{\ell^\prime}^\pm$ is dominant).
We also pointed out that production mechanisms involving the neutral triplet scalars ($H^0$,$A^0$)
of the HTM can contribute to pair production $H^{++}H^{--}$ through the decay chain
$H^0,A^0 \to H^\pm W^*$ followed by  $H^\pm \to H^{\pm\pm}W^*$.
We advocate dedicated simulations of $\qqHpmpmHmp$ with the decay $H^\pm \to H^{\pm\pm}W^*$ (and the
analogous mechanisms with neutral scalars), and a comparison
with $\qqHppHmm$.

\section*{Acknowledgements}
We thank Mayumi Aoki and Koji Tsumura for useful discussions.
 A.G.A was supported by a Marie Curie Incoming International Fellowship, FP7-PEOPLE-2009-IIF, Contract No. 252263.  
 The work of H.S.\ was supported in part
by the Sasakawa Scientific Research Grant
from the Japan Science Society
and Grant-in-Aid for Young Scientists (B)
No.\  23740210.

\end{document}